\documentclass[12pt]{article}
\usepackage{a4}
\usepackage{epsf}
\usepackage{rotating}
\usepackage{epsfig}
\def\Dsl{\,\raise.15ex\hbox{$/$}\mkern-13.5mu D}
\def\beq{\begin{equation}}
\def\eeq{\end{equation}}
\def\bea{\begin{eqnarray}}
\def\eea{\end{eqnarray}}

\newcommand{\vsl}{\slash \kern-12pt\hbox{\it v}}
\newcommand{\sla} [1]{\slash \kern-0.25cm\hbox{\it #1}}

\def\arto{ {\,\,\lower .8ex\hbox {$\longrightarrow
 \atop a \rightarrow 0$}\,\,}}

\def\R1{\varepsilon_1}
\def\E8{\varepsilon_8}

\newcommand{\bd}{\begin{displaymath}}
\newcommand{\ed}{\end{displaymath}}

\newcommand{\be}{\begin{equation}}
\newcommand{\ee}{\end{equation}}
\newcommand{\bi}{\begin{itemize}}
\newcommand{\ei}{\end{itemize}}

\begin{document}
\vspace*{0.5cm}

\setcounter{page}{1}
\begin{flushright}
LPT-ORSAY 02-100\\
FAMN SE-02/02\\
CPHT RR 089.1202\\
UHU-FT/02-24\\
\end{flushright}
\vskip 2.2cm\par
\begin{center}
%\today\hskip 1 cm
%version 2.3

\bf{\huge The strong coupling constant at small momentum   
as an instanton detector. }
\end{center}  
\vskip 0.8cm
\begin{center}{\bf  Ph. Boucaud$^a$, F. De Soto$^{b}$, A. Le Yaouanc$^a$, J.P. Leroy$^a$, 
J. Micheli$^a$, H. Moutarde$^c$,
O. P\`ene$^a$, J. Rodr\'{\i}guez--Quintero$^d$   }\\
\vskip 0.5cm 
$^{a}$ {\sl Laboratoire de Physique Th\'eorique~\footnote{Unit\'e Mixte 
de Recherche du CNRS - UMR 8627}\\
Universit\'e de Paris XI, B\^atiment 210, 91405 Orsay Cedex,
France}\\
$^b${\sl Dpto. de F\'{\i}sica At\'omica, Molecular y Nuclear \\
Universidad de Sevilla, Apdo. 1065, 41080 Sevilla, Spain} \\
$^c$ Centre de Physique Th\'eorique Ecole Polytechnique, 
91128 Palaiseau Cedex, France \\
$^d${\sl Dpto. de F\'{\i}sica Aplicada e Ingenier\'{\i}a el\'ectrica \\
E.P.S. La R\'abida, Universidad de Huelva, 21819 Palos de la fra., Spain} \\
\end{center}
\begin{abstract}
\medskip
We present a study of  $\alpha_{\rm MOM}(p)$ at small $p$ computed from the
lattice. It shows a dramatic  $\propto p^4$ law which 
 can be understood within an instanton liquid model. In this framework 
 the prefactor gives 
 a direct measure of the instanton density in thermalised configurations.   
A preliminary result for this density is $ 5.27(4)\,{\rm fm}^{-4} $.

\noindent P.A.C.S.: 12.38.Aw; 12.38.Gc; 12.38.Cy; 11.15.H
\end{abstract}

\section{Introduction}

In a series of lattice studies~\cite{Boucaud:2000ey}-\cite{Boucaud:2002nc}
the gluon propagator in QCD has been computed at large momenta, and it was 
shown that its behavior was compatible with the perturbative expectation
provided a rather large $1/p^2$ correction was considered. In an OPE approach 
this correction
has been shown~\cite{Boucaud:2000nd,Boucaud:2001st}
to stem from an $A^2$  gluon condensate which has not to vanish
since the calculations are performed in the Landau gauge. 

In the deep infrared (IR) region, where the perturbative approach is completely meaningless, 
the present knowledge of the coupling constant is not so clear, in spite of
all the effort that has been dedicated for years (See \cite{Shirkov:2002gw} and 
references therein).

Lattice calculations provide a unique laboratory to obtain the behavior of 
the coupling at scales below the hadronization scale, where a complete 
understanding of the non-perturbative coupling would be a very important 
step towards the comprehension of hadronization. In this framework, the 
non-perturbative coupling has been computed from different vertices, the 
quark-gluon vertex \cite{Skullerud:2002ge}, 
the ghost-gluon vertex \cite{Alkofer:2000wg}, the three-gluon 
vertex\cite{alles,Boucaud:1998bq}, different propagators 
\cite{Boucaud:1998bq,Nakajima:2002kh}, etc. 

Instantons \cite{thooft} have been proposed, via the instanton liquid 
picture,  to describe a number 
of non-perturbative phenomena in QCD, and particularly to explain 
the low part of the Dirac operator spectrum and hence the  genesis 
of the chiral Goldstone boson, see
 \cite{shuryak}-\cite{hutter} to mention only a few papers.
 
 Instanton studies on the lattice have used essentially 
 the cooled gauge configurations, \cite{teper}-\cite{garciaperez}, which 
 allow to see instanton-like structures. These  results have been used
 to introduce a new definition of the strong coupling constant
 in the IR \cite{Ringwald:1999ze}.

This cooling method has been criticized \cite{Horvath:2002gk}
as creating a distortion on the original thermalised gauge
 configuration, and it was claimed that a direct 
 study of the local chirality on a thermalized gauge configuration
 contradicted the instanton liquid picture. However other authors, 
 \cite{Hip:2001hc,Zhang:rz}, concluded from a similar analysis that 
the dominance of instantons on topological charge fluctuations is not
       ruled out by local chirality measurements. 
       
In this letter we study another observable, namely 
the {\it strong coupling
constant $\alpha_{\rm MOM}$ in the deep IR region on thermalised
gauge configurations} and we show that {\it it strongly supports
the instanton liquid picture}. 

Simultaneously we think that we provide a very simple  and appealing
understanding of the IR behavior of $\alpha_{\rm MOM}$.
This tends to confirm the claim presented in \cite{Boucaud:2002nc}
that an instanton liquid might explain the $<A^2>$ condensate
observed via power corrections to  the perturbative behavior
of the gluon propagator  and $\alpha_{\rm MOM}$ in the 
{\it large momentum regime}.

We will recall the lattice definition of $\alpha_{\rm MOM}$,
derive the  behavior of  $\alpha_{\rm MOM}$
in an instanton liquid and compare the latter with
numerical results in the low momentum region. We then use 
cooled gauge configurations to compare, as a test, the instanton 
density derived 
from $\alpha_{\rm MOM}$ to that which is directly observed
from shape recognition. We then conclude.

%%%%%%%%%%%%%%%%%%%%%%%%%%%%%%%%%%%%%%%%%%%%%%%%%%
\section{Instanton background effect on $\alpha_{\rm MOM}$}
\subsection{Non-perturbative definition of $\alpha_{\rm MOM}$} 
\label{alphanp}

Let us recall shortly the non-perturbative  MOM  definition  of
 $\alpha_s(p^2)$~\cite{alles,Boucaud:1998bq} in Landau gauge.
 We consider the three-gluon Green function 
 ${G^{(3)}}_{\mu_1\mu_2\mu_3}^{a_1 a_2 a_3}(p_1,p_2,p_3)$
 at the symmetric point,
 $p_1^2=p_2^2=p_3^2\equiv \mu^2$.

The tree-diagram three-gluon vertex is given by $g_s\,T^{tree}$ 
with $T^{tree}$ defined by
\beq
 T^{tree}_{\mu_1\mu_2\mu_3}=\big[\delta_{\mu_1'\mu_2'}
 (p_{1}-p_{2})_{\mu_3'}
 + \hbox{cycl. perm.}\big]
 \prod_{i=1,3} \left(\delta_{\mu_i'\mu_i}-\frac {p_{i\,\mu_i'}p_{i\,\mu_i}}
 {p_i^2}\right) \label{tree}
 \eeq 
 The three-gluon Green function may be expanded on a basis
 of tensors. We are interested in the scalar function $G^{(3)}(\mu^2,\mu^2,\mu^2)$
 which multiplies $T^{tree}$. It is obtained by the following contraction
 \[
 G^{(3)}(\mu^2,\mu^2,\mu^2)=\frac {-i} {18 \mu^2}\,\frac{f^{a_1 a_2 a_3}}{24}\,
  {G^{(3)}}_{\mu_1\mu_2\mu_3}^{a_1 a_2 a_3}(p_1,p_2,p_3)\]
 \beq \left[ T^{tree}_{\mu_1\mu_2\mu_3} + \frac{(p_1-p_2)_{\mu_3}
(p_2-p_3)_{\mu_1}(p_3-p_1)_{\mu_2}}{2 \mu^2}\right]\label{projsym}
 \eeq 
 
 The Euclidean two point Green function in momentum space writes in the
 Landau Gauge: 
\beq\label{prop}
        {G^{(2)}}_{\mu_1\mu_2}^{a_1 a_2}(p,-p)=G^{(2)}(p^2) 
        \delta_{a_1 a_2} \left(\delta_{\mu_1\mu_2}-
        \frac{p_{\mu_1}p_{\mu_2}}{p^2}\right)\label{G2}
\eeq
where $a_1, a_2$ are the color indices ranging from 1 to 8. 

Then the renormalised coupling constant is 
 given by
 \cite{alles} 
 \beq
 g_R(\mu^2)= \frac{G^{(3)}(p_1^2,p_2^2,p_3^2) Z_3^{3/2}(\mu^2)}
 {G^{(2)}(p_1^2)G^{(2)}(p_2^2)G^{(2)}(p_3^2)}\label{gr}
 \eeq 
where 
\beq
        Z_3(\mu^2)= G^{(2)}(\mu^2) \mu^2\label{Z3}
\eeq

\subsection{Solution in an instanton liquid } 

Let us consider an instanton liquid picture~\cite{shuryak}-\cite{hutter}
i.e. a gauge field given by

\beq
A_\mu^{(I) a}({\bf x})=\sum_i R^{ a \alpha}_{(i)}\, \overline{\eta}^\alpha_{\mu
\nu}{(x_\nu-z_{\nu}^i)} \ \rho_i^{-2}
P\left(\frac{(x-z^i)^2}{\rho_i^2}\right) \ ,
\label{amuins}
\eeq

\noindent where $z^i$ ($\rho_i$) are the center (radius) of the instantons,
$\overline{\eta}^\alpha_{\mu \nu}$ is known as 't Hooft symbol,
 $R^{ a\, \alpha}_{(i)}$
are color rotations embedding the canonical  $SU(2)$ instanton into the $SU(3)$
gauge group, $\alpha=1,3$ ($ a=1,8$) is an $SU(2)$ ($SU(3)$)  color index, and the
sum is extended over instantons and anti-instantons.

In order to take into account instanton deformation resulting from their 
interaction~\footnote{We assume a non-negligible  instanton
density.} we {\it do not} assume that the radial function  $P(u)$
is equal to $2/(u^2(u^2+1))$ as for 't Hooft-Polyakov's instanton.

The  field's Fourier transform is

\bea
\label{Ak}
\widetilde{A_\mu^{(I) a}}({\bf p})  =i \sum_i R^{ a \alpha}_{(i)}
\overline{\eta}^\alpha_{\mu \nu} e^{ik\cdot z^i}\rho_i^3\frac{p_\nu}{p}
I(p \rho_i)
\eea

\noindent for   $p\ne  0$ and  $\widetilde{A_\mu^{(I) a}}(0)=0$.
The function $I(p\rho_i)$ is given by
\beq
I(s) \ = \frac{4 \pi^2} s\int_0^\infty\ z^3 dz  J_2(sz)\,P(z) \ ;
\label{Igen}
\eeq
$J_2$ being the  second order Bessel J~function.

From eqs.~(\ref{prop}), (\ref{Ak}) and the relation   
\beq
{G^{(2)}}_{\mu \nu}^{ a b}(p)=\frac 1 V <\widetilde{A_\mu^{ a}}({\bf p})
\widetilde{A_\mu^{b}}({\bf -p})>
\eeq 
where the average $<\cdots>$ stands here for the average over instantons, 
we get
\beq\label{propins}
G^{(2)}(p^2)=  \frac n 8 <\rho^6  I(p\rho)^2>
\eeq
$n$ being the  instanton density  and $\rho$ being the instanton radius.
When computing  the  result in eq. (\ref{propins}) we have used 
the relations concerning the 
 ${\eta}$ tensors from the appendix A in ref.~\cite{shuryak} and the
orthogonality relation 
\beq\label{2points}
\sum_ a R^{ a \alpha}_{(i)}R^{ a \beta}_{(i)}  = \delta^{\alpha,\beta}\qquad
 <\sum_ a R^{ a \alpha}_{(i)}R^{ a \beta}_{(j)} e^{ip(z_i-z_j)}>_{i\ne j} 
 \simeq 0 \,\,
\eeq 
For brevity we skip the derivation of the first of these equations.
The second one is not exact but expresses the
hypothesis~\footnote{In ref. \cite{diakonov}, using variational methods it 
is shown that the color orientation of different instantons seems
 to be weakly correlated.} that
the color rotations are at random for different  instantons as well as 
the phases $e^{ip(z_i-z_j)}$. The cross-products are thus reasonably assumed to
 average to zero.

Similarly, 
\beq
G^{(3)}(p^2,p^2,p^2)= \frac n {48\, p}<\rho^9  I(p\rho)^3>
\eeq
where we have used again ref.~\cite{shuryak} and
\beq
\sum_{ a b c} R^{ a \alpha}_{(i)}R^{b \beta}_{(i)}R^{c \gamma}_{(i)}
f_{ a b c} = \epsilon^{\alpha\beta\gamma}.
\eeq
We skip again the derivation of this equation 
and again we neglect crossed terms for the same reason  as in (\ref{2points}).
The strong coupling constant is then given by
\beq
\alpha_{\rm  MOM}(p)  \equiv \frac {g_R^2}{4\pi} = \frac 1{4\pi}
\left[\frac{G^{(3)}(p^2,p^2,p^2)}{(G^{(2)}(p^2))^3}(p^2G^{(2)}(p^2))^{3/2}\right]^2
=\frac 1  {18 \pi} n^{-1} p^4
\label{amom}
\eeq
where  we have assumed, for simplicity, that all the instantons have equal
 radii; this will be discussed later.
The very remarkable feature of this result is that {\it it does  not depend
on the shape of the instanton-like structures} i.e. on the function $I$ 
(that has all the information about the profile function $P(u)$), 
neither on the scale  $\rho$ appearing in (\ref{amuins}) i.e.  on the instanton
radius. {\it It only depends 
on the instanton density $n$.}

\section{Lattice three gluon vertex.}

\begin{figure}[ht] 
\begin{center} 
%\begin{tabular}{c c} 
%\includegraphics[width=16pc]{pert_alpha.eps}  & 
%\includegraphics[width=16pc]{fit_all.eps} \\ 
%  a & b \\ 
%\end{tabular} 
\includegraphics[width=32pc]{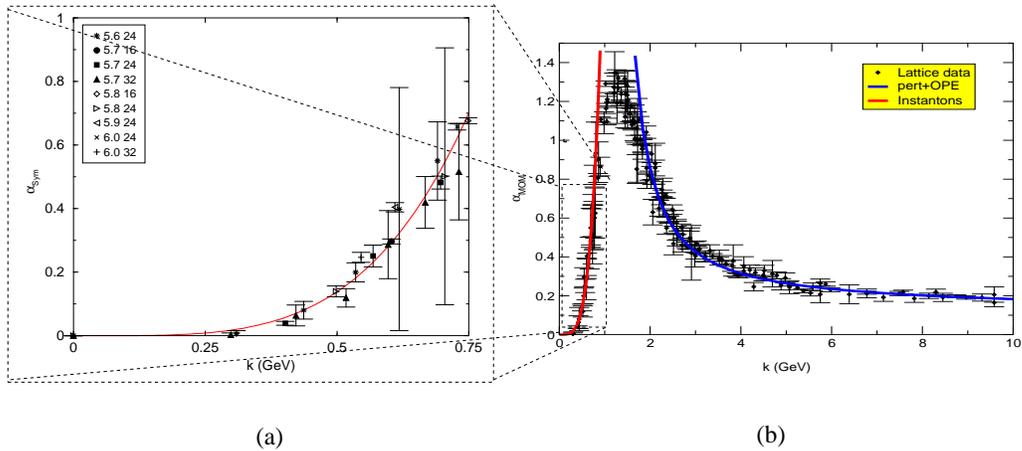} 
\caption{\small 
(b) Symmetric MOM coupling constant for different lattice settings 
and fits to perturbative expression plus power corrections in the high 
momenta region and to expression (\ref{amom}) discussed in the text 
for small momenta. (a) Region of small momenta is zoomed.} 
\label{alpha_sym} 
\end{center} 
\end{figure}

In several lattice works, the scheme outlined in \ref{alphanp} has been 
used \cite{Boucaud:1998bq} to compute the running coupling constant in a 
quite wide range of lattice volumes and spacings. The region of high momenta
has been precisely described according to the perturbative running plus 
non-perturbative corrections in an Operator Product Expansion 
framework (Figure \ref{alpha_sym}(b)). 

The smaller momenta,  on the contrary, are not yet theoretically understood. 
 The aim of this paper is to try an instanton interpretation
 at very low momentum, as far as possible from the perturbative regime.
 We will therefore make a fit of the deep IR region
combining all points from different lattice  settings in order to achieve enough
statistics. The use of different lattice settings, with varying statistics, makes 
it difficult to estimate the errors with the usual jacknife method.
 Therefore, calling $\chi_{\rm min}^2$ the minimum $\chi^2$,  we 
estimate the errors by assuming that one standard deviation is reached when
$\chi^2=\chi_{\rm min}^2+1$ while, of course, the central value
 corresponds to
$\chi^2=\chi_{\rm min}^2$.

A quick inspection of the points in fig. \ref{alpha_sym}(a)
shows that even for very small momenta the scaling is very good, i.e.
the simulations with different lattice spacings agree strikingly.
We have used unusually low values of $\beta$ i.e. unusually large 
lattice spacings - and hence large volumes - in order to reach 
small momenta. The quality of the scaling for these small 
momenta as well as for larger ones makes us confident  
that we do  not introduce a sizable bias with  these small $\beta$'s.
We take the ratios between lattice spacings for different $\beta$'s from
 ref.~\cite{Guagnelli:2000jw} and $a^{-1}(\beta=6.0)= 1.97$ GeV.

If we fit the low tail of the coupling (up to momenta
$\sim 0.8\ {\rm GeV}$) by a power law, a power $3.84(8)$
is obtained, in good agreement with the expected power-four 
behavior. 

Fixing now the power to be four, the density resulting from
formula (\ref{amom}) is $n = 5.27(4){\rm fm}^{-4}$, 
with $\chi^2/d.o.f.=3.8$ (Figure \ref{alpha_sym}(a)).
This is in the right ballpark since several arguments~\cite{Negele:1998ev}
point toward a few  ${\rm fm}^{-4}$'s.
The rather large $\chi^2/d.o.f.=3.8$ is due to two points which have 
unusually small errors, maybe because of the small statistics used
in this preliminary study. Furthermore an exact power 4 should not be taken 
too seriously since the instanton radii distribution can presumably
distort this power law. Indeed the dispersion of the 't Hooft
instanton radii leads to an effective power smaller than four at
small momenta, but the deviation from four is small.

%Furthermore an exact power 4 should not be taken 
%too seriously since the instanton radii distribution could presumably
%distort this power law. However, we studied this effect for a
%distribution of 't Hooft instantons and found an effective power 
%slightly smaller than four.

So within this approach, {\it we are able to compute the instanton density
directly from the thermalised lattice, and we obtain a result which
is not biased by a cooling procedure}.

%%%%%%%%%%%%%%%%%%%%%%%%%%%%%%%%%%%%%%%%%%%%%%%%%%%%%%%%%%%%%%%%%%%%% 
%%%%%%%%%%%%%%%%%%%%%%%%%%%%%%%%%%%%%%%%%%%%%%%%%%%%%%%%%%%%%%%%%%%%%%

An important question arises here. How can we be sure 
that what we see are really instanton-like objects ?
We have already noticed that the result in (\ref{amom}) is 
obtained for any radial profile of the semi-classical structures
 considered, {\it provided the tensorial structure is that
of} eq. (\ref{amuins}). But it is easy to see that any 
 semi-classical field configuration different from eq. (\ref{amuins})
 would also produce a $p^4$ law but {\it with a different prefactor}.
 Therefore we might as well have seen other structures than instantons
 and thus our estimate of the density, which relies on the prefactor 
 as compared to the one-instanton prediction, could be wrong.
 
\begin{figure}[ht]
\begin{center}
\begin{tabular}{c c c}
\includegraphics[width=14.5pc]{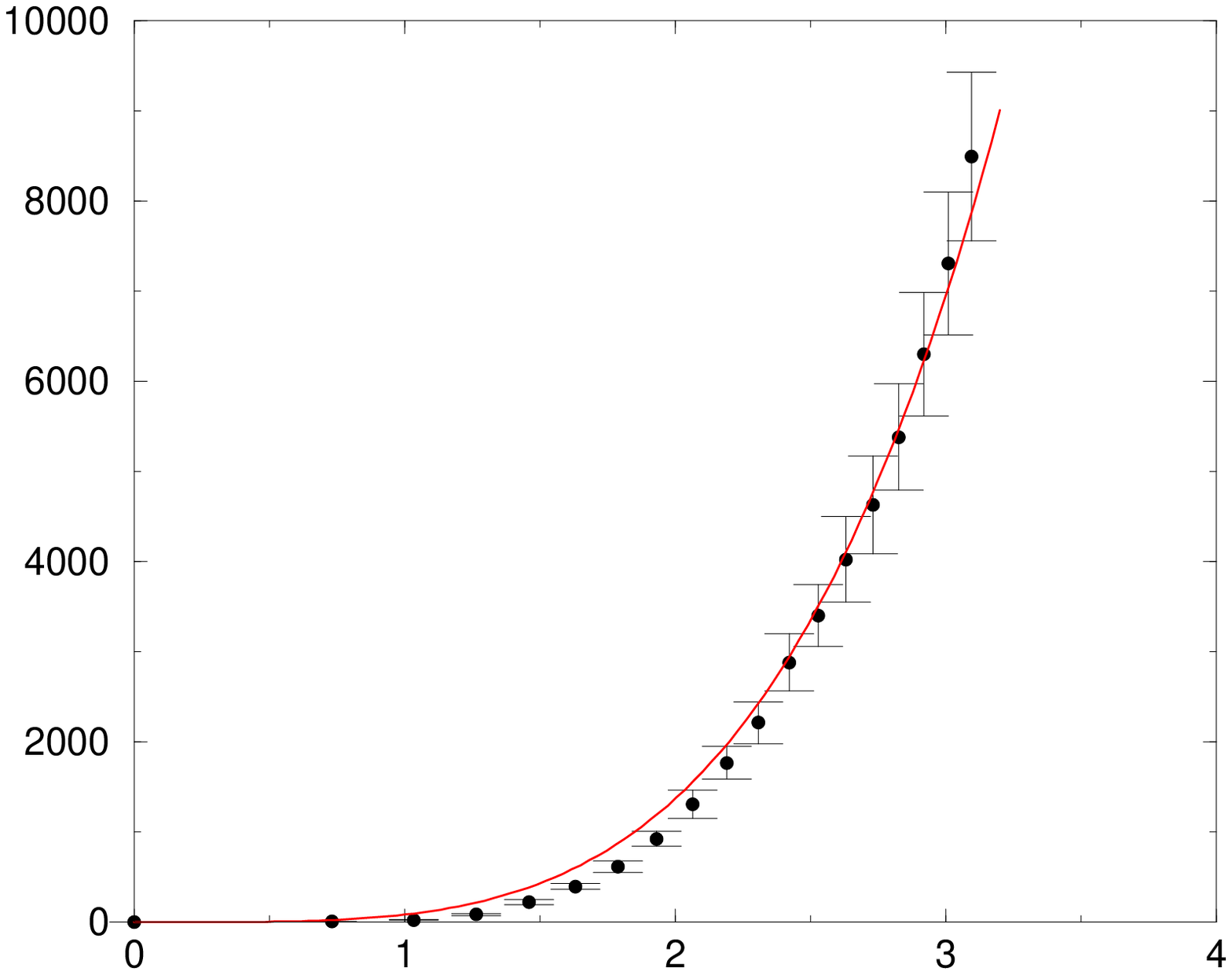}  & 
\hskip 0.5 cm\includegraphics[width=16.5pc]{alpha_plot_log2.eps} \\
  a &    b \\
\end{tabular}
\caption{\small (a) Coupling constant in a cooled lattice ($L=24$,$\beta=6.0$) 
after 200 cooling sweeps. The solid line corresponds to the fit discussed in 
the text. The horizontal axis is given in GeV, assuming for simplicity
the lattice spacing of the thermalised configurations, 
$a^{-1}=1.97$ GeV.
 (b) Coupling constant in a cooled lattice ($L=24$,$\beta=5.6$) 
after several  cooling sweeps. The horizontal axis is
the momentum assuming the same lattice spacing, $a^{-1}=0.83$ GeV, which is the 
value for the thermalised configurations.
This is a log-log plot which exhibits
better the power law. There seems to be three regimes. }
\label{alpha_sym_cool}
\end{center}
\end{figure}

 In order to check our interpretation we appeal to cooled configurations.
 We insist that we do  not use the   controversial cooled configurations
 to infer the properties  of the thermalised ones, but only to check
 our  new method to estimate the instanton density against the 
 instanton shape recognition (ISR) method which can only be applied after
 cooling. 
 We cool the configurations according to  the method described 
 in~\cite{Boucaud:2002nc} and compute the 
 ``strong coupling constant''~\footnote{For simplicity we call
 ``coupling constant'' this quantity computed 
according to the definitions in section \ref{alphanp} although 
being aware that in  cooled configurations this denomination
is not really appropriate.} 
 according to the scheme described section \ref{alphanp}. 
 The first striking result is that after cooling 
the coupling constant {\it grows in 
the whole energy range} (Figure \ref{alpha_sym_cool}).   It reaches 
exceedingly large values  at large  momenta, the  reason  of  which 
being the $\propto p^4$ law and  the large prefactor due
to a small instanton density, see eq. (\ref{amom}).

  We now make a fit to a $p^4$ law for $\alpha_{\rm MOM}(p)$  
  on lattice configurations which have been submitted to a large number
 of cooling sweeps (200) so that UV fluctuations 
should have disappeared leaving only semiclassical structures.
We do not take in the fit the few smallest momenta for reasons which 
will be discussed soon. We  compare the instanton density
extracted via formula (\ref{amom}) applied to the  $\propto  p^4$
fit of $\alpha_{\rm MOM}(p)$ to the instanton density  coming from 
the geometrical ``instanton shape recognition'' (ISR) method described 
in \cite{Boucaud:2002nc}. We obtain 
the qualitative agreement shown in table \ref{compISR}. Notice that these
values for the instanton densities have nothing 
to do with the latter density in the thermalised configuration~\footnote{For
simplicity we have used the lattice spacing of the thermalised configurations.}.
What is relevant is the fair agreement between both methods  to 
 estimate the instanton density.

\begin{table}[hbt]
\begin{center}
\begin{tabular}{|c c c c|}
\hline
L  &$\beta$ &  n(ISR) &   n($\alpha$) \\
\hline
24 &5.6     &0.009(6)   &0.019(1)\\
24 &5.8     &0.048(23)  &0.090(1)\\
24 &6.0     &0.115(6)   &0.133(2)\\
32 &6.0     &0.145(16)  &0.197(7)\\
\hline
\end{tabular}
\caption{\small Comparison between the instanton density, n (in ${\rm fm}^{-4}$), 
obtained through the Instanton Shape Recognition (ISR) method 
\cite{Boucaud:2002nc} and the density deduced from the fit of the 
coupling constant after 200 cooling sweeps. In both cases errors 
are only statistical.}
\label{compISR}
\end{center}
\end{table}

We expect differences in table \ref{compISR} between both estimates,
 mainly at small $\beta$, because the ISR method does not 
  recognize small instantons in lattice units, which induces
for ISR a systematic underestimate
   of the density \cite{addendumA2}. The general tendency in table \ref{compISR}
supports this argument, as both estimates of the density are
closer for large values of beta~\footnote{As the lattice spacing is increased
an increasingly large number of instantons are missed by the ISR method because
they become too small in lattice units.}.

In figure \ref{alpha_sym_cool}(a) it can be seen that at low momentum the
points for $\alpha_{\rm MOM}$ are below the fit. The log-log plot of the
same quantity for different cooling sweeps in  figure \ref{alpha_sym_cool}(b)
confirms that the behavior at small momenta differs slightly from the large
momenta one. In fact the detailed dependence of the slopes in the momentum and
the number of cooling sweeps seems to be involved but it is striking
that the slope  stays always in the range three to five, i.e. close to
the expected slope four.
%We have studied the effect of the dispersion of the instanton radii
%$\rho_i$.  It leads to an effective power smaller than four  at small
%momenta but the deviation from four is small. 

Anyhow, why do the thermal configuration seem to agree at small momentum 
 with the $\propto  p^4$ law, fig.  \ref{alpha_sym}(a), while the cooled
 configurations seem unexpectedly to agree with the same law only  at larger
 momentum~? One might argue that at
large distance instanton deformation~\cite{addendumA2}
 as well as instanton
(anti-)instanton repulsion (attraction)  might  have been generated by
the cooling itself~\footnote{During  the cooling instanton
anti-instanton pair annihilation occurs which confirms  that
correlations are generated by the cooling.}. This would explain why the
$p^4$ law observed at small momentum in the thermalised configurations
tends to be distorted with cooling.  
Honestly this argument  has to be submitted to closer scrutiny
and for  now we  consider the detailed understanding of these slopes
as an open question, but we would like to stress that anyhow {\it the observed
behaviour is never far from $\propto  p^4$}.

%%%%%%%%%%%%%%%%%%%%%%%%%%%%%%%%%%%%%%%%%%%%%%%%%%%%%%%%%%%%%%%%%%%%%
%%%%%%%%%%%%%%%%%%%%%%%%%%%%%%%%%%%%%%%%%%%%%%%%%%%%%%%%%%%%%%%%%%%%%%

\section{Conclusion and discussion}

\begin{itemize}
\item We have found that the lattice simulations indicate good  scaling of 
$\alpha_{\rm MOM}(p)$  when the lattice spacing is varied, 
even at rather small $\beta$ when $p$ is small.
\item  We find 
\beq
\alpha_{\rm MOM}(p) \simeq \frac 1  {18 \pi} n^{-1} p^4, \quad {\rm 
for}\quad p\le 0.8 {\rm  GeV} \eeq 
as expected from  an instanton liquid
picture and  eq. (\ref{amom}).
\item The fitted  density  
\beq\label{dens} n
= 5.27(4){\rm fm}^{-4} \eeq
 is in fair agreement with expectations.

\item
We have checked that the calculation 
of the instanton density from eq. (\ref{amom}) via a $\propto p^4$
fit of $\alpha_{\rm MOM}(p)$ is comparable to a direct counting
of instantons recognised from their shape. 
We take this as a convincing evidence that this new 
 $\alpha_{\rm MOM}$-method to measure instanton density
is reliable. 

\end{itemize}

{\it This makes it highly plausible that this $\alpha_{\rm MOM}(p)\propto p^4$
is an effect of a liquid  of instantons, i.e.
that such a liquid of  instantons indeed exists  in the thermalised
configurations and that the quantum fluctuations do not affect
significantly  $\alpha_{\rm MOM}(p)$ for \,$p\le 0.8$ GeV.   }

These results open a Pandora box of new questions: can this
simple explanation also apply to other definitions of $\alpha_s$, 
to other Green functions (the gluon  propagator in particular) ? 
How is this interpretation related  to Schwinger-Dyson or 
renormalisation group deduced small momentum behavior of 
these quantities~?

A look at figure \ref{alpha_sym} (b) shows a nice theoretical 
understanding (solid lines) of $\alpha_{\rm MOM}(p)$ both in the large 
(perturbative QCD + OPE) and small (instanton liquid picture) 
momentum regimes.  
How to understand better the transition between these
two regimes ?

\section{Acknowledgments} 
 We are grateful to  Dmitri Shirkov for illuminating
 discussions. 
This work was supported in part by the European Network 
"Hadron Phenomenology from
Lattice QCD" HPRN-CT-2000-00145 and by Picasso agreement HF2000-0056. 
F.S. is indebted to the 
Spanish Fundaci\'on C\'amara 
for financial support.

\end{document}